\documentclass[runningheads]{llncs}
\usepackage[T1]{fontenc}

\usepackage{graphicx}

\usepackage{amsfonts}
\usepackage{amssymb, amsmath}
\usepackage{bm}
\usepackage{subcaption}

\begin{document}

\title{Quantifying Feature Importance of Games and Strategies via Shapley Values}

\author{Satoru Fujii\inst{1}} 

\authorrunning{S. Fujii}

\institute{Kyoto University, Yoshida-honmachi Kyoto, Japan \email{fujii.satoru.75c@st.kyoto-u.ac.jp}} 

\maketitle           

\begin{abstract}
Recent advances in game informatics have enabled us to find strong strategies across a diverse range of games. However, these strategies are usually difficult for humans to interpret. On the other hand, research in Explainable Artificial Intelligence (XAI) has seen a notable surge in scholarly activity. Interpreting strong or near-optimal strategies or the game itself can provide valuable insights. In this paper, we propose two methods to quantify the feature importance using Shapley values: one for the game itself and another for individual AIs. We empirically show that our proposed methods yield intuitive explanations that resonate with and augment human understanding.

\end{abstract}

\section{Introduction}

Recent advancements in machine learning technology have facilitated the analysis of increasingly large games with greater precision. Notably, the amalgamation of conventional algorithms with deep neural networks — exemplified by AlphaZero \cite{silver2018general} in perfect information games, Deep CFR \cite{brown2019deep} and NFSP \cite{heinrich2016deep} in imperfect information games, has demonstrated outstanding performance, occasionally surpassing top human players.

While the development of high-performing AI for games is an intriguing objective, our primary interest lies in deepening our own understanding of the game. In recent years, there has been a rise in professional players endeavoring to enhance their strategies by studying the moves of AI in domains such as shogi. The trend of humans assimilating AI's decision-making processes, or collaborating with AI to reach decisions, is anticipated to proliferate, extending beyond board games to diverse fields. However, these AIs inherently lack transparency. Due to their vastness and intricacy, their strategies elude human comprehension. This absence of explainability poses challenges.

In realms like data analytics and image processing, there is burgeoning interest in eXplainable AI (XAI) — a branch that renders machine learning outcomes interpretable to humans. Yet, in the gaming domain, such research has not been sufficiently conducted. At present, human players either unconditionally adopt strategies recommended by advanced AI or resort to observational learning. Introducing explainability to game AI would streamline the process of grasping the AI's thoughts. In addition, explaining AI strategies that are sufficiently close to optimal ones can be considered as characterizing the essence of the game itself, facilitating a profound understanding of the game's nature.

In this paper, we propose two Shapley-based approaches to achieve this goal. Firstly, we introduce a method to quantify the feature importance of the game itself by utilizing the expected return of abstracted games with obscured features. This approach elucidates which features are crucial to act optimally. Secondly, we present a model-agnostic method for calculating the future importance of individual AIs without requiring additional training. This clarifies features that the AI has attention to when choosing actions. 

Our empirical findings not only validate their theoretical basis but also reveal that they resonate with and augment human's perceptions of games and strategies.

\section{Notation and Background}

\subsection{Extensive-form Games}
Many games can be formally represented as extensive-form games, comprising the following elements:
\begin{enumerate}
\item $\mathcal{N} = (1,\cdots,n)$ denotes the set of players involved in the game.
\item Game tree $T$ is a rooted tree that encapsulates the state transition rules of the game. $Z$ denotes the set of all terminal nodes. Non-terminal nodes are referred to as \textit{turns} and are denoted by $X$. The outgoing edges originating from a node $x \in X$ are termed \textit{actions} and are denoted as $A(x)$. Since nodes in game tree contain information regarding all preceding actions, these nodes are referred to as the \textit{history}.
\item $\mathcal{P}=(\mathcal{P}_0,\mathcal{P}_1,\cdots,\mathcal{P}_n)$ constitutes a partition of $X$ and signifies the turns for players $1,\cdots,n$. The nodes denoted by $\mathcal{P}_0$ represent \textit{chance} events, where the actions are determined by a probability distribution function $p: \mathcal{P}_0 \to A(\mathcal{P}_0)$.
\item $\mathcal{I}_1,\cdots,\mathcal{I}_n$ form a subpartition of $\mathcal{P}_1,\cdots,\mathcal{P}_n$. Each $I \in \mathcal{I}_i$ is called \textit{infoset}, wherein all nodes in $I$ are indistinguishable to player $i$. Nodes within the same infoset are required to have an identical set of actions, which can consequently be denoted as $A(I)$.
\item A utility function $u_i: Z \to \mathbb{R}$ yields the payoff for player $i \in \mathcal{N}$ at terminal nodes. If $\sum_{i \in \mathcal{N}} u_i(z) = 0$ for all $z \in Z$, the game is categorized as a \textit{zero-sum} game.
\end{enumerate}

If all infosets of every player consist of a single node, the game is called \textit{perfect information}; otherwise, it is a game of \textit{imperfect information}.

For each player $i$ and for each of their infoset $I \in \mathcal{I}_i$, \textit{behavior strategy} $\sigma_i$ maps each action $a \in A(I)$ to its selection probability. The set of all possible behavior strategies for player $i$ s denoted by $\Sigma_i$.

A strategy profile $\sigma$ refers to the tuple of behavior strategies for all players. Additionally, $\sigma_{-i}$ denotes the set of strategies from $\sigma$ excluding $\sigma_i$.

When players act according to a strategy profile $\sigma$, the probability of reaching a history $h$ and an infoset $I$ are written as $\pi^{\sigma}(h)$, $\pi^{\sigma}(I)$ respectively. The \textit{expected payoff} for player $i$ under strategy profile $\sigma$ is defined as:
\[
u_i(\sigma) = \sum_{h \in Z}\pi^{\sigma}(h)u_i(h)
\]

\subsection{Exploitability}

Let $b_i(\sigma_{-i})$ denote the best-response payoff of player $i$ against the other players' strategies $\sigma_{-i}$, which is formally defined as:
\[
b_i(\sigma_{-i}) = \max_{\sigma'_i \in \Sigma_i} u_i(\sigma'_i, \sigma_{-i},)
\]
Here, $\sigma'_i \in \Sigma_i$ that realizes this maximum is termed as the \textit{best response}. Various methods exist to compute the best-response payoff, such as transforming the problem into a Linear Programming (LP) \cite{10.5555/1296179}.

A strategy profile $\sigma^* = (\sigma^*_1,\cdots,\sigma^*_n)$ is said to be a \textit{Nash equilibrium} if, for all players $i = 1,\cdots,n$ and for all possible strategies $\sigma_i$, the following inequality holds:
\[
u_i(\sigma^*) \geq u_i(\sigma_i, \sigma^*_{-i})
\]
It is known that at least one such Nash equilibrium exists.

In two-player zero-sum game, \textit{exploitability} of $\sigma_1, \sigma_2$ is respectively defined as
\[
\epsilon_1(\sigma_1) = b_2(\sigma_1) - v_2^*
\]
\[
\epsilon_2(\sigma_2) = b_1(\sigma_2) - v_1^*
\]
where $v_i^*$ denotes the expected payoff for player $i$ in a Nash equilibrium.

\subsection{Counterfactual Regret Minimization (CFR)}

Counterfactual Regret Minimization (CFR) \cite{zinkevich2007regret} is an iterative algorithm designed to approximate a Nash equilibrium on imperfect information games. During each traversal of the game tree, it computes values termed \textit{counterfactual regrets} for all infosets and subsequently updates its strategy based on the average counterfactual regret accumulated over the traversals.

In two-player zero-sum games, it is established that the average of such strategies converges to a Nash equilibrium. However, the computational complexity of full game tree traversal often renders such methods impractical for large games. As a result, Monte Carlo CFR (MCCFR) \cite{lanctot2009monte} is generally preferred, as it samples actions and traverses only a portion of the game tree, thereby providing a more computationally efficient alternative. Deep CFR \cite{brown2019deep} addresses larger games by using neural networks to approximate average counterfactual regret and average strategy, which eliminates the necessity of storing those values in memory.

\subsection{Abstraction}

Given the difficulty of applying learning algorithms to intractably large games, \textit{abstraciton} is commonly employed to turn them into smaller-sized games. Formally, an abstraction of infosets $\alpha_i$ is defined as a subpartition of $\mathcal{P}_i$ that is coarser than $\mathcal{I}_i$. Let $\Gamma^{\alpha_i}$ be the extensive-form game derived by substituting $I_i$ with $\alpha_i$ in the original extensive-form game $\Gamma$, and $\sigma^{*, \alpha_i}_i$ be a strategy of player $i$ in a Nash equilibrim of $\Gamma^{\alpha_i}$. It is important to note that infosets of only a single player are abstracted in this case. In two-player zero-sum games, the following theorem has been proven \cite{waugh2009abstraction}:

\begin{theorem} \label{mono}
If abstraction $\alpha'_i$ is a subpartion of $\alpha_i$, 
\[
\epsilon_i(\sigma^{*, \alpha_i}_i) \geq \epsilon_i(\sigma^{*, \alpha'_i}_i) 
\]
\end{theorem}

This theorem means that an equilibrium strategy in finer abstraction is less exploitable in the original game than one in coarser abstraction, ensuring the theoretical soundness of using abstraction. In addition, it is also pointed out that abstracted player's strategy in this equilibrium is the least exploitable strategy that can be represented in the space \cite{johanson2012finding}. This theorem does not hold when infosets of the other player are also abstracted.

\subsection{Feature Importance and Shapley Value}


\textit{Additive feature attribution} \cite{lundberg2017unified} is a prevalent technique for interpreting machine learning models by assigning importance to each feature of individual inputs. Here, we assume that a model $f$ has an input vector $\bm{x}$ comprising $m$ features. Further, we assume that $f$ can produce an output even if some of these features are invisible. For $\bm{z}' \in \{0,1\} ^m$, we will use $h_{\bm{x}} (\bm{z}')$ to denote the input derived by obscuring features that have 0 value in $\bm{z}'$ in the original input $\bm{x}$. Let $\bm{x}' \in \{0,1\}^M$ be the vector that all features visible in $\bm{x}$ are 1. This implies $h_{\bm{x}}(\bm{x}') = \bm{x}$.

In additive feature attribution, an \textit{explanation model} $g$, an interpretable local approximation of the original model around $\bm{x}$, is written as:
\[
g(\bm{z}') = \phi_0 + \sum_{i=1}^m \phi_i z'_i
\]
Here, $\phi_0$ is the default output when all features are invisible, and $\phi_i$ quantifies the contribution of $i$-th feature to the output value of $f(\bm{x})$.

Let $\phi_i(f,\bm{x})$ be a function that yields $\phi_j$ from $f$ and $\bm{x}$. Additionally, $\bm{z}' \setminus j$ represents the vector $\bm{z}'$ except its $j$-th element set to 0. The following properties are desired for the explanation model $g$.

\begin{enumerate}
\item Local Accuracy: $f(\bm{x}) = g(\bm{x}') = \phi_0 + \sum_{j=1}^m \phi_j x'_j$
\item Missingness: $x'_j = 0 \Rightarrow \phi_j = 0$
\item Consistency: For any two models $f$ and $f'$, if
\[f'(h_{\bm{x}}(\bm{z}')) - f'(h_{\bm{x}}(\bm{z}' \setminus j)) \geq f(h_{\bm{x}}(\bm{z}')) - f(h_{\bm{x}}(\bm{z}' \setminus j))\]
for all $\bm{z}' \in \{0,1\}^m$, then $\phi_j(f', \bm{x}) \geq \phi_j(f,\bm{x})$.
\end{enumerate}

The following additive feature attribution, known as \textit{Shapley additive explanation} (SHAP) \cite{lundberg2017unified}, is verified to be the sole additive feature attribution that satisfies all of these properties:
\[
\phi_j(f,\bm{x}) = \sum_{\bm{z}' \subseteq \bm{x}'} \frac{|\bm{z}'|! (m - |\bm{z}'| - 1)!}{m!} (f(h_{\bm{x}}(\bm{z}')) - f(h_{\bm{x}}(\bm{z}' \setminus j)))
\]
In the preceding formula, $|\bm{z}'|$ is the number of non-zero elements in $\bm{z}$, and $\bm{z}' \subseteq \bm{x}'$ means $\bm{z}'$ has non-zero elements that constitute a subset of those in $\bm{x}'$. This value is equivalent to the average marginal contributions of feature $j$ across all permutations. Given the challenges of calculating the exact SHAP value, random sampling approaches \cite{vstrumbelj2014explaining} are often preferred. For neural network models, Deep SHAP \cite{lundberg2017unified} offers an efficient computation alternative.

While SHAP focuses on explaining local predictions of the model, there also has been a study using Shapley values for assessing global feature importance, essentially averaging the marginal contributions to guarantee similar desirable properties. \cite{covert2020understanding}.

\section{Related Work}

In the field of Reinforcement Learning (RL) for single agent environments, a significant number of research focused on explainability have been conducted \cite{heuillet2021explainability}. There is a recent work \cite{beechey2023explaining} delving into Shapley-based feature importance in RL using value functions and policies. Another notable contribution \cite{greydanus2018visualizing} employs a saliency map of the game screen, highlighting the agent's focus and effectively indicating the feature importance of each pixel.

For cooperative multi-agent environments, various studies \cite{han2022stable,li2021shapley} have utilized Shapley values to allocate the credit or the reward amongst agents.

\section{Descriptions of Our Methods}

We introduce two general approaches to quantify feature importance for different subjects: the first seeks to provide a global interpretation of the game itself, while the second aims to elucidate local strategies employed by individual AIs. We focus on two-player zero-sum games. For the target player of explanation $i$, we postulate that each infoset $I \in \mathcal{I}_i$ can be represented by $m$ features. We let $\mathcal{M}$ be the set of $m$ features. Formally, we assume that feature function $F_j: \mathcal{I}_i \rightarrow \mathcal{F}_j$ is given for each $j \in \mathcal{M}$, where each $\mathcal{F}_j$ is a finite set of values.

\subsection{Shapley Game Feature Importance}

To quantify the global feature importance of the game for the target player $i$, we compute an approximated Nash equilibrium $\hat{\sigma}^{*,\alpha}$ of games with abstraction $\alpha$ that some of $m$ features are obscured to player $i$ while the opponent plays in null abstraction to hold Theorem \ref{mono}, then calculate the expected return of $i$. 

For a set of visible features $S \in 2^\mathcal{M}$, we define $\alpha_i(S)$ as the abstraction that merges infosets sharing the identical set of actions and feature values $F_j(I)$ for each $j \in S$. The difference between $u_i(\sigma^{*, \alpha_i(S)})$ and $u_i(\sigma^{*, \alpha_i(\mathcal{M})})$ quantifies the importance of invisible features $\mathcal{M} - S$. This can subsequently be employed to calculate Shapley feature importance by averaging marginal contributions across all possible feature permutations, which we call \textit{Shapley Game Feature Importance (SGFI)}.

\subsection{Shapley Strategy Feature Importance}

To quantify the feature importance of $\sigma_i(I)$ for a given $\sigma_i$, we use the sampled SHAP value \cite{vstrumbelj2014explaining} with modifications for games, calculated by the following algorithm:

Firstly, we sample $I^{\text{rnd}} \sim \{I' \in \mathcal{I}_i \ | \ A(I') = A(I)\}$ $t_1$ times, then assign the average of $\sigma_i(I^{\text{rnd}})$ to $\phi_0$. Then, we repeat the following steps $t_2$ times to calculate $\phi_j$ for each feature $j$ by averaging the output $\phi'_j$.
\begin{enumerate}
\item Sample $I^{\text{alt}} \sim \{I' \in \mathcal{I}_i \ | \ A(I') = A(I)\}$.
\item Randomly choose $\mathcal{O}$ from $m!$ possible permutations of $m$ features.
\item Sample $I^{\text{s}} \sim \{I' \in \mathcal{I}_i \ | \ F_p(I') = F_p(I) , \ F_q(I') = F_q(I^{\text{alt}}),  \ \text{for all} \ p \in P \cup \{j\} \ \text{and} \ q \in Q\}$.
\item Sample $I^{\text{s,alt}} \sim \{I' \in \mathcal{I}_i \ | \ F_p(I') = F_p(I) , \ F_q(I') = F_q(I^{\text{alt}}),  \ \text{for all} \ p \in P \ \text{and} \ q \in Q \cup \{j\}\}$.
\item $\phi'_j \gets \sigma_i(I^\text{s}) - \sigma_i(I^\text{s,alt})$
\end{enumerate}
In step $3.$ and $4.$, $P$ denotes the set of the features ahead of $j$ in $\mathcal{O}$ and $Q$ denotes $\mathcal{M} - P - \{j\}$. If the corresponding infoset does not exist, we assign $\phi_0$ for $\sigma_i(I^\text{s})$ or $\sigma_i(I^\text{s,alt})$. We call $(\phi_1, \cdots, \phi_m)$ computed by this algorithm as \textit{Shapley Strategy Feature Importance (SSFI)}.

\section{Experiments} \label{experiments}

\subsection{Goofspiel}

We used Goofspiel as the subject of the experiment. The adopted rule set in this study is delineated as follows:

The game is played between two players. Each player starts with a hand of $k$ distinct cards, labeled from 1 to $k$. Alongside the players' hands, an identical deck of $k$ cards is shuffled to constitute a draw pile.

The gameplay is divided into $k$ rounds. During each round, a single card is randomly drawn from the draw pile and placed face-up at the center of the table. Subsequently, both players simultaneously select and reveal one card from their respective remaining hands. The player who revealed the card with the higher number receives points equal to the number of the center card. In the case of a tie, neither player is awarded any points. Cards utilized in each round are then excluded from further play. The player accumulating the highest total points across all rounds is declared the winner.

While Goofspiel is designed as a simultaneous-move game, it can be reformulated as a sequential-move game without altering its core mechanics, by allowing one player to select their card first, unbeknownst to the opposing player.

We arranged the following 4 features to explain each infoset of Goofspiel:
\begin{enumerate}
    \item Center (C) : The center card 
    \item Deck (D) : Cards remaining in the draw pile
    \item Opponent (O) : Cards remaining in the opponent's hand
    \item Point (P) : The difference of points between the target player and the opponent
\end{enumerate}
We denote the corresponding feature functions as $F_C, F_D, F_O, F_P$. This entails $\mathcal{F}_C = \{1, \cdots, k\}$, $\mathcal{F}_D = \mathcal{F}_O = 2^{\{1, \cdots, k\}}$, $\mathcal{F}_P = \{-k(k+1)/2, \cdots, k(k+1)/2\}$. Note that cards remaining in the player's hand constitute the action set and are always visible to the player.

\subsection{SGFI of Goofspiel}

\begin{figure}[h]
  \def\@captype{table}
  \begin{minipage}[t]{.48\textwidth}
    \begin{center}
\begin{tabular}{|l|r|r|} \hline
  & SGFI \\ \hline
Center & 0.298  \\
Deck & 0.295 \\
Opponent & 0.096  \\
Point & 0.101 \\ \hline
\end{tabular}
    \end{center}
    \caption{SGFI of Goofspiel}
    \label{sgfi}
  \end{minipage}
  \hfill
  \begin{minipage}[c]{.48\textwidth}
 \centering
 \includegraphics[keepaspectratio, scale=0.26, bb=0 0 623 606]
      {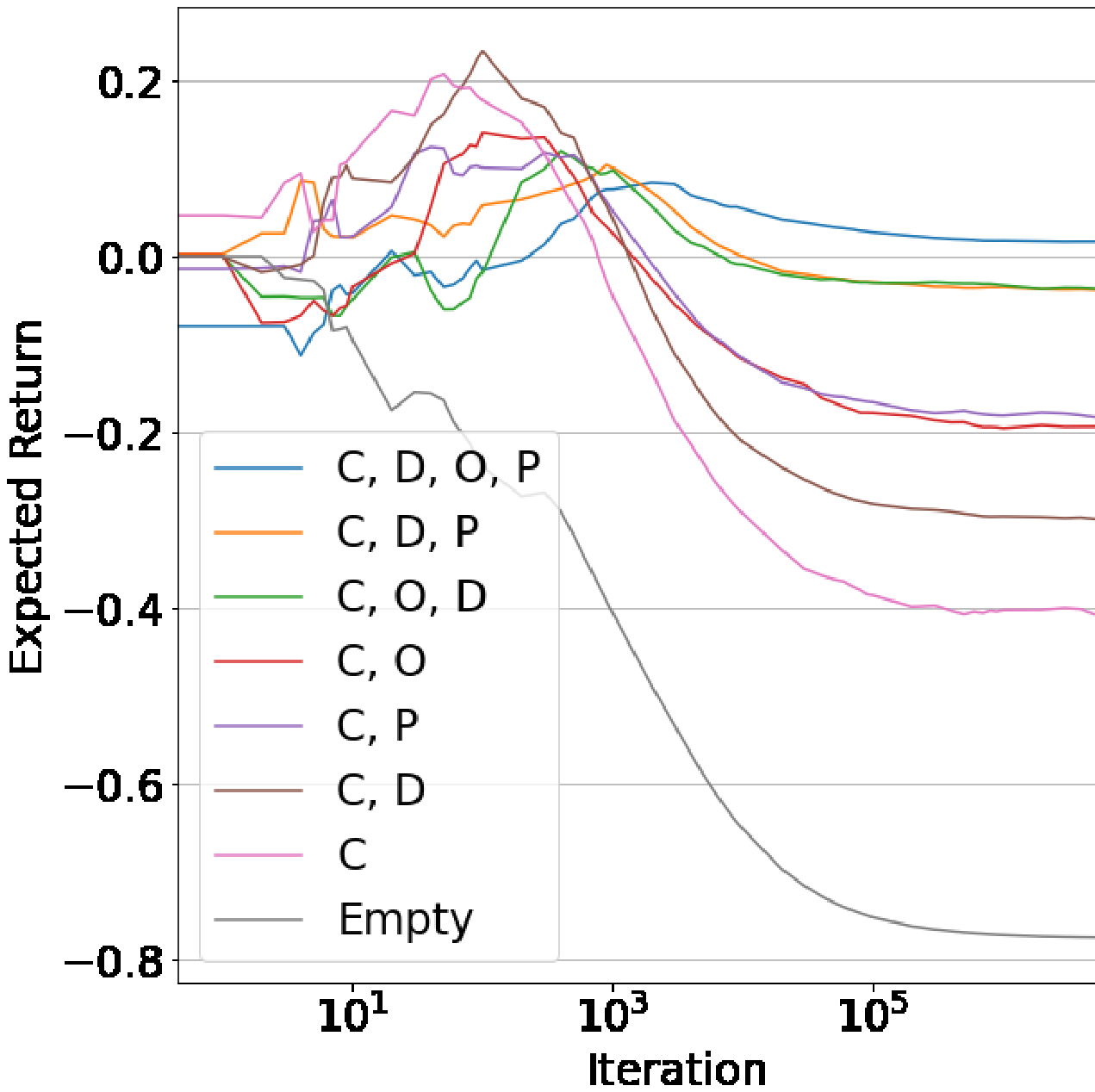}
    \caption{Expected return convergence of the target player for different abstractions. Legends denote elements in $S$.}
    \label{ret1}
  \end{minipage}
\end{figure}

\begin{figure*}
\begin{tabular}{cc}
\begin{minipage}[b]{0.48\linewidth}
 \centering
 \includegraphics[keepaspectratio, scale=0.26, bb=0 0 623 606]
      {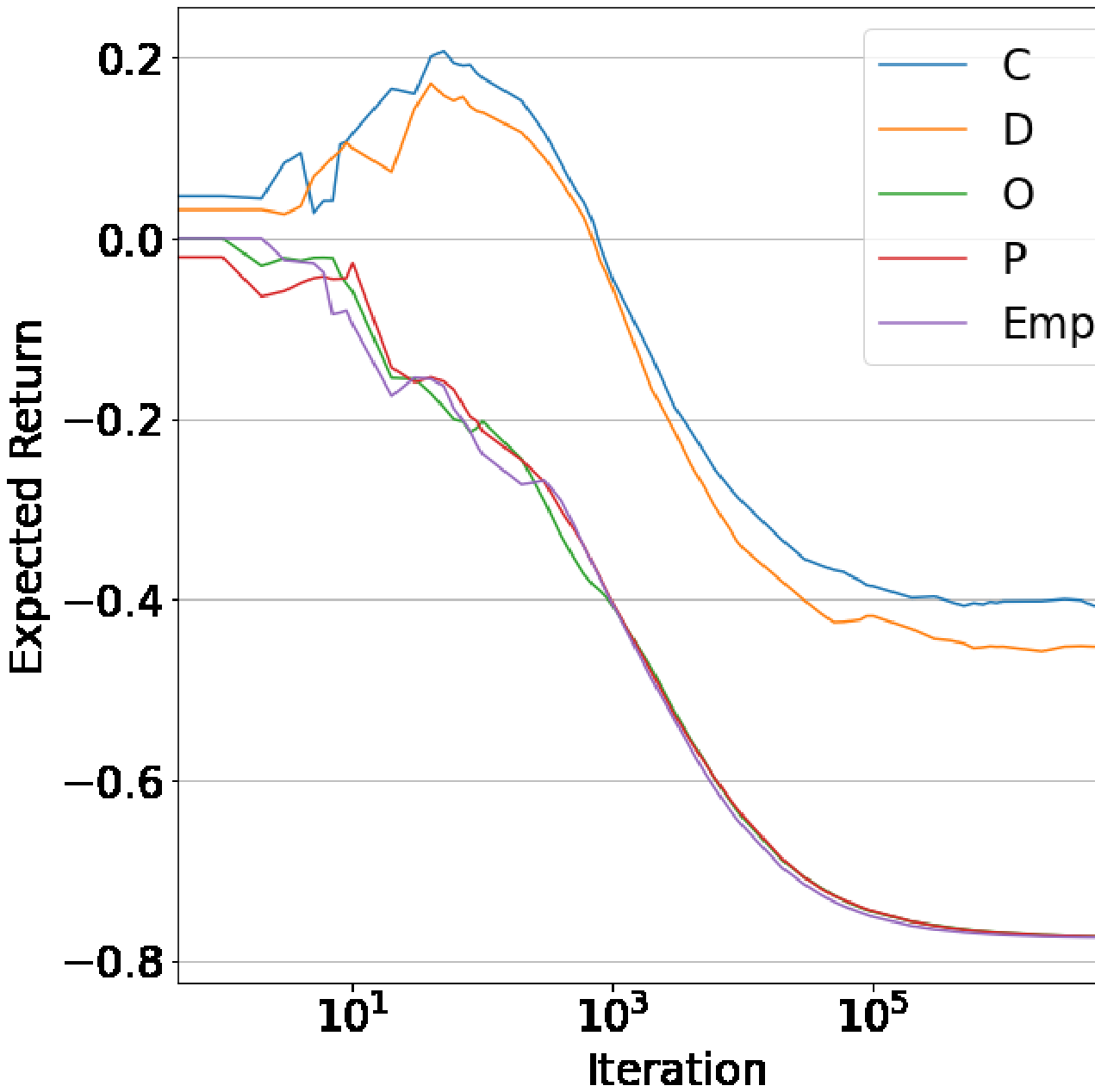}
\end{minipage} &
\begin{minipage}[b]{0.48\linewidth}
 \centering
 \includegraphics[keepaspectratio, scale=0.26, bb=0 0 623 606]
      {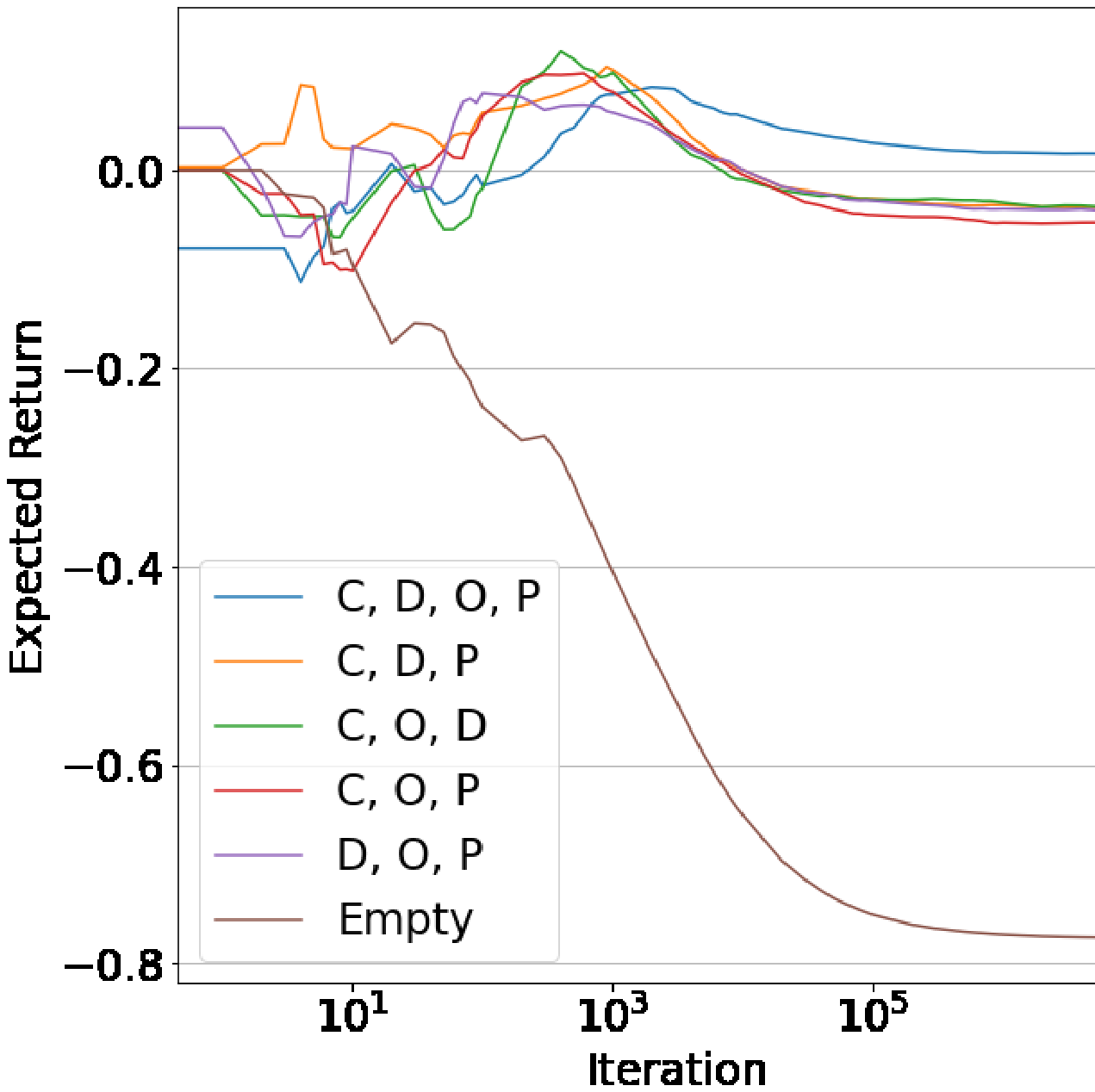}
\end{minipage}
\end{tabular}
\caption{Convergence of $u_i(\hat{\sigma}_i^{*, \alpha_i(\{j\})})$ and $u_i(\hat{\sigma}_i^{*, \alpha_i(\mathcal{M}-\{j\})})$. Legends denote elements in $S$.} \label{ret2}
\end{figure*}

We employed external sampling MCCFR \cite{lanctot2009monte} with $10^7$ timesteps to approximate a Nash equilibrium strategy for all $2^4$ patterns of abstracted Goofspiels in $k=5$, using OpenSpiel \cite{LanctotEtAl2019OpenSpiel} as the learning environment. CFR-BR \cite{johanson2012finding} could be employed for larger games where learning with a null abstraction player is infeasible.

Fig. \ref{sgfi} shows the resulting SGFI values of the four features. Given that the most intuitive strategies in Goofspiel involve selecting a high card when a high center card is present and conserving high cards when they remain in the deck, it is intuitive that the feature importances of both the Center and Deck surpass the others.

Fig. \ref{ret1} depicts the convergence of expected return across various abstractions. It can be seen that adding features increases the expected return. It is also noteworthy that when $S=\mathcal{M}(=\{C,D,O,P\})$, the expected return of the target player ended up slightly positive because this abstraction contains sufficient information while a null abstraction suffers slight redundancy, resulting in the target player's faster learning. Fig. \ref{ret2} shows the convergence of $u_i(\hat{\sigma}^{*, \alpha_i(\{j\})})$ and $u_i(\hat{\sigma}^{*, \alpha_i(\mathcal{M}-\{j\})})$. It can be seen that solely using Opponent or Point does not markedly impact the performance in comparison to $S=\varnothing$, while the combination of these features with the others can significantly influence the outcomes as shown in Fig. \ref{ret1}. It is also noticeable that obscuring a single feature yields similar results regardless of the obscured one. These results underscores the benefits of employing the Shapley value as an index instead of simply using $u_i(\hat{\sigma}^{*, \alpha_i(\{j\})}) - u_i(\hat{\sigma}^{*, \alpha_i(\emptyset)})$ or $u_i(\hat{\sigma}^{*, \alpha_i(\mathcal{M})}) - u_i(\hat{\sigma}^{*, \alpha_i(\mathcal{M}-\{j\})})$ as the feature importance of $j \in \mathcal{M}$.

\subsection{SSFI of Goofspiel AI}

We once again utilized external sampling MCCFR with $10^6$ timesteps for Goofspiel in $k=4$ without any abstraction in order to obtain the AI to explain. This resulted in strategy profile $\hat{\sigma*}$ exhibiting an average exploitability of $0.006$. In SSFI calculations, both $t_1$ and $t_2$ were set to $10^6$. 

\begin{figure}[h]
  \begin{minipage}[c]{.48\textwidth}
 \centering
 \includegraphics[keepaspectratio, scale=0.23, bb=0 0 623 606]
      {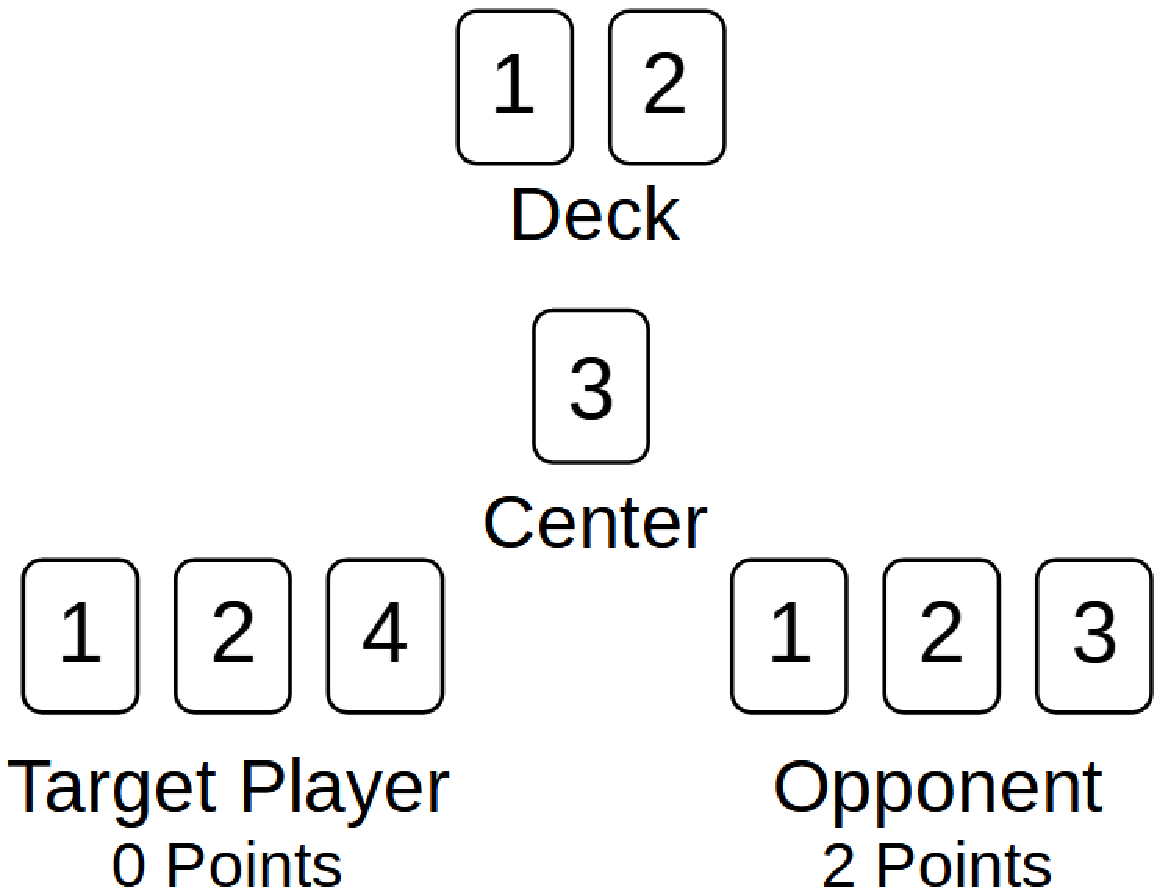}
    \caption{Situation in $I^1$}
    \label{i1}
  \end{minipage}
  \hfill
  \def\@captype{table}
  \begin{minipage}[c]{0.48\hsize}
    \centering
\begin{tabular}{|l|rrr|}
\hline
 & Card 1     & Card 2     & Card 4    \\ \hline
$\phi_0$   & 32.5\%  & 35.3\%  & 32.2\% \\
$\phi_C$ & -16.3\% & +16.1\% & +0.2\% \\
$\phi_D$ & -9.6\%  & +16.9\% & -7.3\% \\
$\phi_O$ & -6.5\%  & +12.7\% & -6.2\% \\ \hline
$\hat{\sigma*}(I^1)$ & 0.1\%   & 81.0\%  & 18.9\% \\ \hline
\end{tabular}
\caption{SSFI of $\hat{\sigma}^*_1(I^1)$.} 
    \label{table_a}
  \end{minipage}
\end{figure}

Table \ref{table_a} shows the SSFI values of $\hat{\sigma}^*_1(I^1)$, where $A(I^1) = \{1,2,4\}, \ F_C(I^1)=3, \ F_D(I^1)=\{1,4\}, \ F_O(I^1)=\{1,2,3\}, \ F_P(I^1)=-2$. $F_P$ was omitted from the SSFI calculation since the value of Point ($-2$) can be derived from $A(I^1)$ and the other features. It can be seen that $\phi_0 + \phi_C + \phi_D + \phi_O = \hat{\sigma}^*_1(I^1)$: starting from the default strategy $\phi_0$, each $\phi_j$ elucidates the contribution of feature $j$ to the strategy. This renders the explanation more comprehensible.

\begin{figure}[h]
  \begin{minipage}[c]{.48\textwidth}
 \centering
 \includegraphics[keepaspectratio, scale=0.23, bb=0 0 623 606]
      {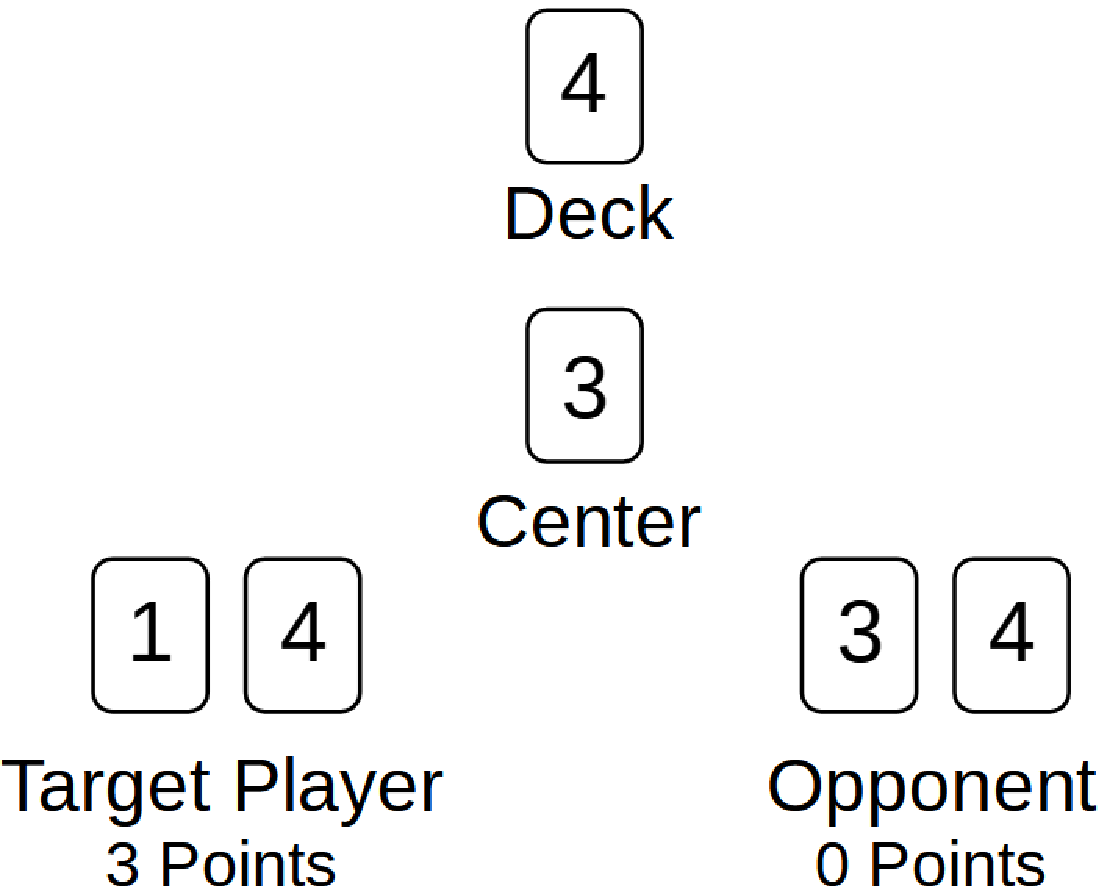}
    \caption{Situation in $I^2$}
    \label{i2}
  \end{minipage}
  \hfill
  \def\@captype{table}
  \begin{minipage}[c]{0.48\hsize}
    \centering
\begin{tabular}{|l|rr|}
\hline
 & Card 1 & Card 4    \\ \hline
$\phi_0$   & 49.5\%                     & 50.5\%          \\
$\phi_C$   & +5.4\%                     & -5.4\%          \\
$\phi_D$   & +4.5\%                     & -4.5\%          \\
$\phi_O$   & +3.8\%                     & -3.8\%          \\
$\phi_P$  & +4.2\%                     & -4.2\% \\ \hline
$\hat{\sigma*}(I^2)$ & 67.0\%                     & 33.0\%          \\ \hline
\end{tabular}
\caption{SSFI of $\hat{\sigma}^*_1(I^2)$.} 
    \label{table_b}
  \end{minipage}
\end{figure}

Table \ref{table_b} gives another example of SSFI, where $A(I^2)=\{1,4\}, F_C(I^2)=3, \ F_D(I^2)=\{4\}, \ F_O(I^2)=\{3,4\}, \ F_P(I^2)=3$ in this target infoset $I^2$. It is noticeable that the values across features in this scenario are more congruent, indicating that the significance of a feature might oscillate between different infosets.

\section{Conclusion}

In this paper, we have put forward two methods for the quantification of feature importance, each with its distinct application: understanding the game on a macro level and deciphering the attention of individual AI strategies. Other than providing the theoretical backgrounds of those methods, we have empirically demonstrated that our methods can explain games and strategies in a way that aligns with and complements human intuition.

We need to keep in mind that XAIs using feature importance have inherent limitations in that they solely teach you what feature they regard as important. However, we posit that such feature importance can enhance our understanding of more intricate strategies and provide better heuristics for the game.

We anticipate that interpretative methodologies, such as those we have proposed, will enhance human learning supported by AI and streamline collaborative decision-making processes between humans and AI.

\section*{Acknowledgement}
I would like to thank Professor Hideki Tsuiki for meaningful discussions. I am also grateful to the ACG referees for useful comments.

\bibliographystyle{splncs04}
\bibliography{biblo}

\end{document}